\documentclass[prd,preprint,tightenlines,superscriptaddress,floatfix,showpacs,preprintnumbers,nofootinbib,eqsecnum]{revtex4}

 \usepackage[dvips,final]{graphicx}
     \usepackage{epsfig}
      \usepackage{bm}

\begin{document}

\thispagestyle{empty} \preprint{\hbox{}} \vspace*{-10mm}

\title{Exclusive $D \bar D$ meson pair production\\
in peripheral ultrarelativistic heavy ion collisions}

\author{M.~{\L}uszczak}
\email{luszczak@univ.rzeszow.pl}
\affiliation{University of Rzesz\'ow, PL-35-959 Rzesz\'ow, Poland}

\author{A.~Szczurek}
\email{antoni.szczurek@ifj.edu.pl}

\affiliation{Institute of Nuclear Physics PAN, PL-31-342 Cracow, Poland} 
\affiliation{University of Rzesz\'ow, PL-35-959 Rzesz\'ow, Poland}

\date{\today}

\begin{abstract}
The cross sections for exclusive $D^+D^-$ and $D^0 \bar {D^0}$ meson 
pair production in peripheral nucleus - nucleus collisions are calculated and
several differential distributions are presented.
The calculation of the elementary  $\gamma \gamma \to D \bar D$
cross section is done within the heavy-quark approximation and in the Brodsky- Lapage formalism with distribution
amplitudes describing recent CLEO data on leptonic $D^+$ decay.
Realistic (Fourier transform of charge density)
charge form factors of nuclei are used to generate photon flux factors.
Absorption effects are discussed and quantified.
The cross sections of a few nb are predicted for RHIC and of a few hundreds of nb for LHC
with details depending on the approximation made in calculating elementary $\gamma \gamma \to D \bar D$ cross sections.
\end{abstract}
\pacs{25.75.-q,25.75.Dw,25.20Lj}

\maketitle

\section{Introduction}

The main aim of the heavy ion program
at ultrarelativistic collision energies is concentrated on the discovery
and  analysis of the quark-gluon plasma.
High charges of the colliding ions give a possibility to
study also peripheral processes when only a few particles are
produced. In particular, large fluxes of photons
associated with the huge charges of nuclei open an interesting
possibility to study photon-photon collisions \cite{Baur} which are
difficult to study in $e^+ e^-$ and proton-proton collisons.

In our earlier works we have studied a production of $\mu^+ \mu^-$ \cite{KS2010},
$\rho^0 \rho^0$ \cite{KS2009} and recently of $c \bar c$ and $b \bar b$ \cite{KSMS2010}. We have demonstrated how important is the inclusion of the realistic charge
form factors responsible for generating photon flux factors.
In our analysis charge form factors of nuclei are calculated as Fourier
transform of the realistic charge densities as measured in electron scattering
off nuclei.

Many years ago Brodsky and Lepage have suggested how to calculate
large-angle production of light meson pairs ($\pi \pi$, $K \bar K$) in photon- photon
collisions.
In the meantime several further studies have been preformed and some 
improvements have been suggested. 
Parallel those processes were searched for in several experimental studies.

Heavy quark meson pair production was studied theoretically only in Ref.\cite{Choi1994} where
formulas have been derived in the heavy quark approximation with Dirac delta-like distribution amplitudes.
On the other hand both lattice QCD \cite{D_DA_lattice} and the CLEO collaboration \cite{D_DA_experiment} extracted the D-meson distribution amplitude which turned out to differ considerably from the delta-like
distribution amplitude assumed in heavy quark approximation. In the present studies we will use also the more
realistic distribution amplitudes.

Both LEP2 and Belle studies were not able to extract corresponding cross sections. 
The nuclear processes look potentially interesting in this respect due to the
large charges of the colliding ions.
In the present analysis we wish to calculate cross sections for exclusive production of 
$D^+D^-$ meson pairs in $A A \to A A D \bar D$ 
reactions at RHIC (Au+Au, $W_{NN}$ = 200 GeV) and
LHC (Pb+Pb, $W_{NN}$ = 5.5 TeV). Whether the experimental analysis is possible requires experimental feasibility studies.

\begin{figure}[!h]    
\begin{minipage}[t]{0.4\textwidth}
\centering
\includegraphics[width=0.9\textwidth]{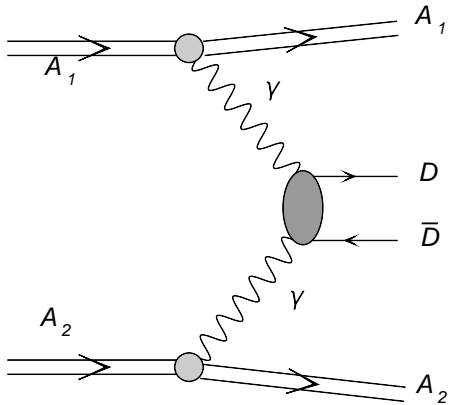}
\end{minipage}
   \caption{\label{fig:diagram1}
   \small The diagram for the exclusive $D \bar D$ meson production.
}
\end{figure}

In Fig.\ref{fig:diagram1} we show the basic QED mechanism
of the exclusive production of $D \bar D$ pairs in the peripheral heavy-ion collisions.
We consider both $D^+D^-$ and $D^0 \bar{D^0}$ production.

In the next section we concentrate on the calculation of the cross section 
for the $\gamma \gamma \to D \bar D$ reaction which, as discussed in section
\ref{sec:gamma_gamma}, is used in the equivalent photon approximation
in the impact parameter space (b-space EPA). The nuclear cross sections are presented
in the result section. Conclusions and summary close our paper.

\section{Heavy charmed meson pair production in photon- photon collisions}

\label{sec:gamma_gamma}

\begin{figure}[!h]   
\includegraphics[width=0.12\textwidth]{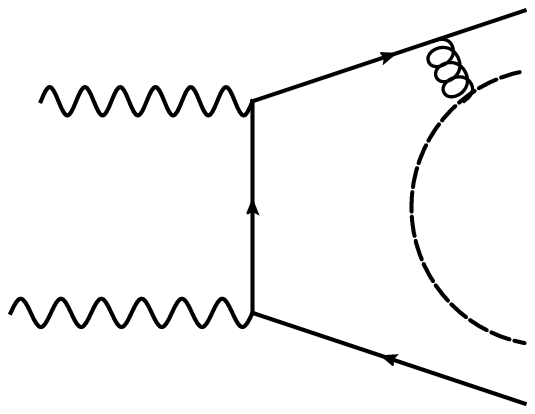}
\includegraphics[width=0.12\textwidth]{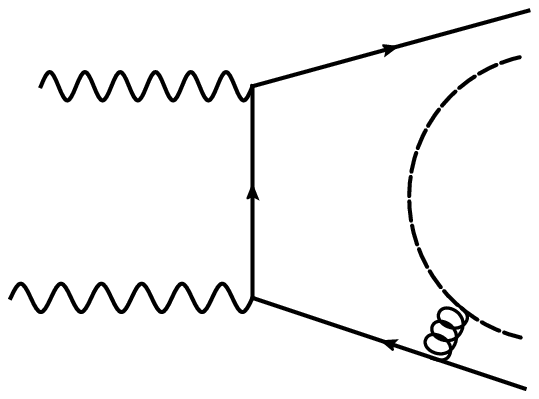}
\includegraphics[width=0.12\textwidth]{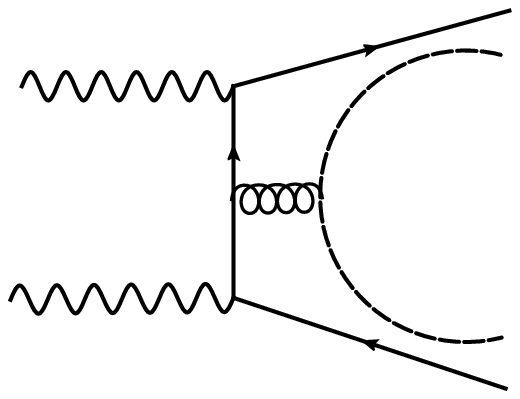}\\

\includegraphics[width=0.12\textwidth]{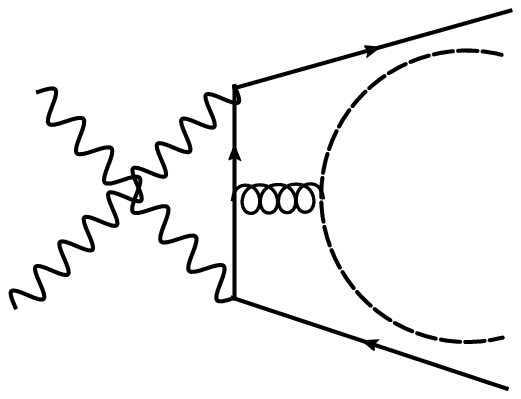}
\includegraphics[width=0.12\textwidth]{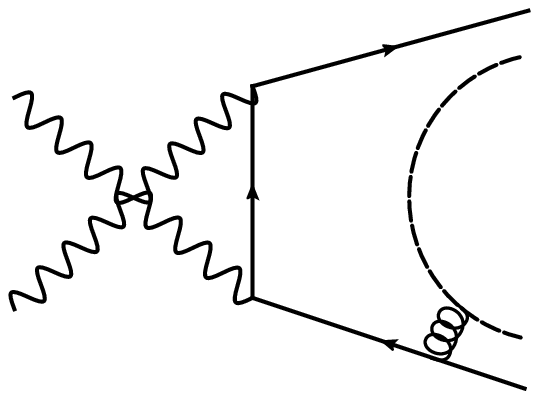}
\includegraphics[width=0.12\textwidth]{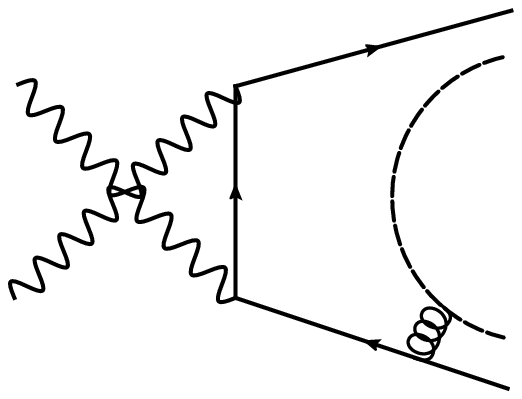}\\

\includegraphics[width=0.12\textwidth]{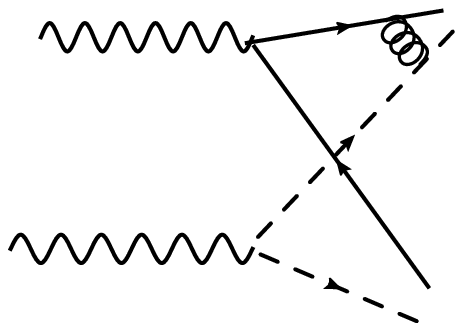}
\includegraphics[width=0.12\textwidth]{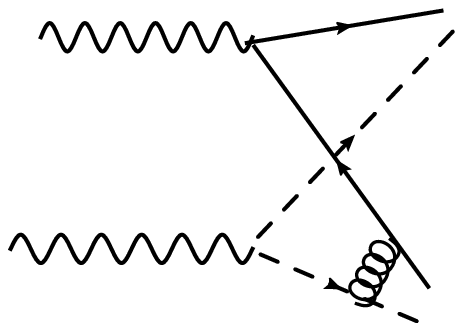}
\includegraphics[width=0.12\textwidth]{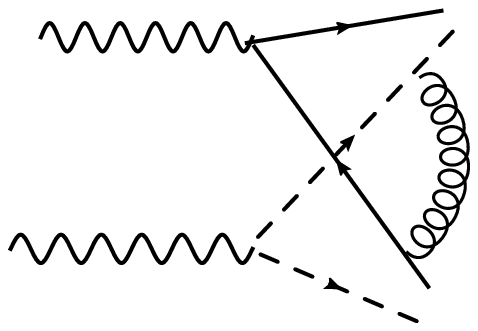}\\

\includegraphics[width=0.12\textwidth]{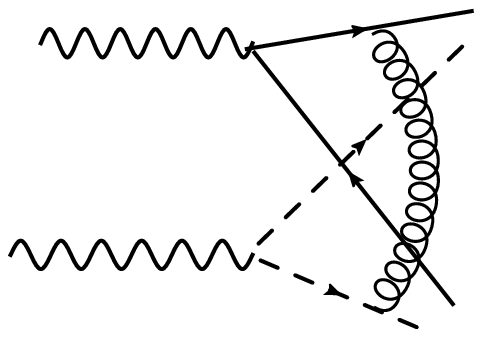}
\includegraphics[width=0.12\textwidth]{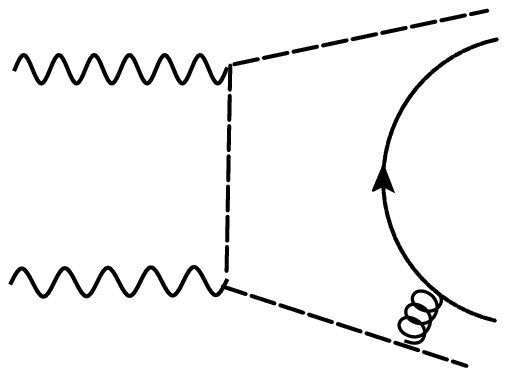}
\includegraphics[width=0.12\textwidth]{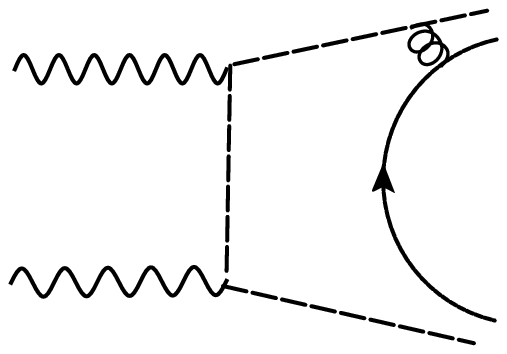}\\

\includegraphics[width=0.12\textwidth]{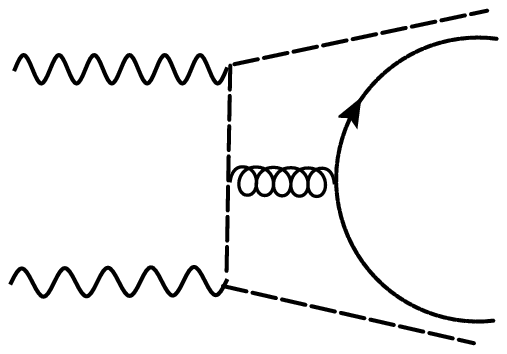}
\includegraphics[width=0.12\textwidth]{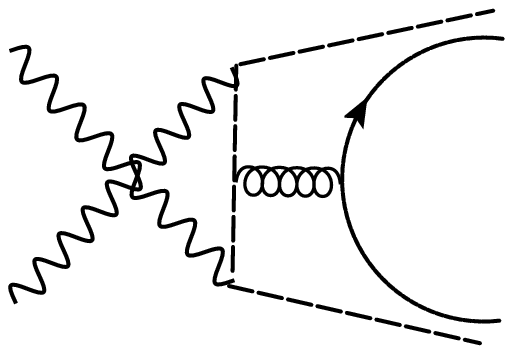}
\includegraphics[width=0.12\textwidth]{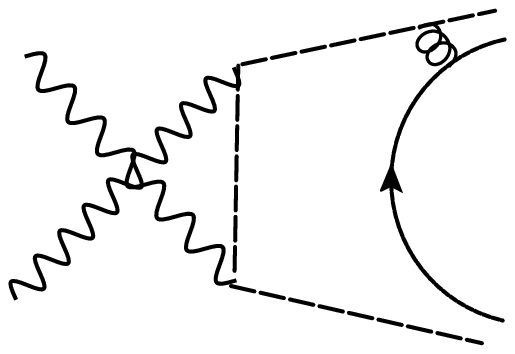}
\includegraphics[width=0.12\textwidth]{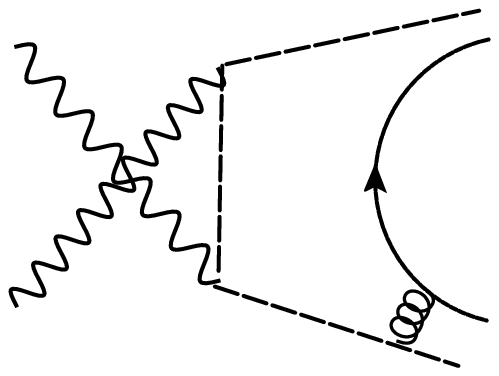}\\

\includegraphics[width=0.12\textwidth]{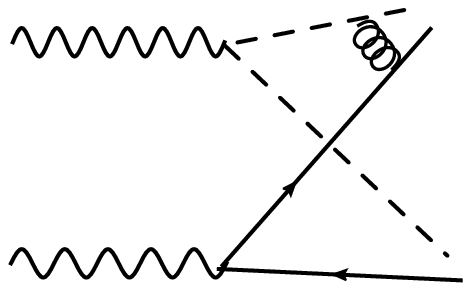}
\includegraphics[width=0.12\textwidth]{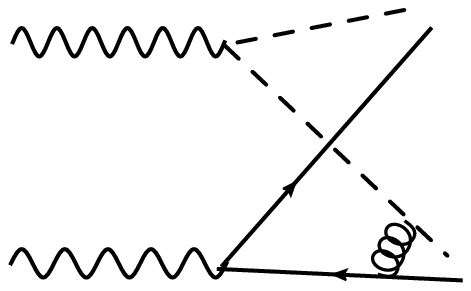}
\includegraphics[width=0.12\textwidth]{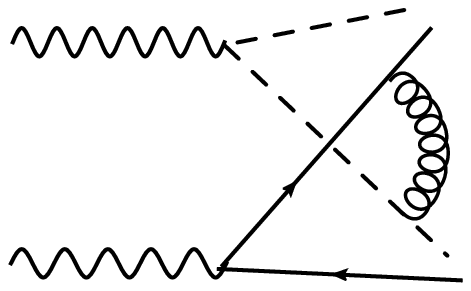}
\includegraphics[width=0.12\textwidth]{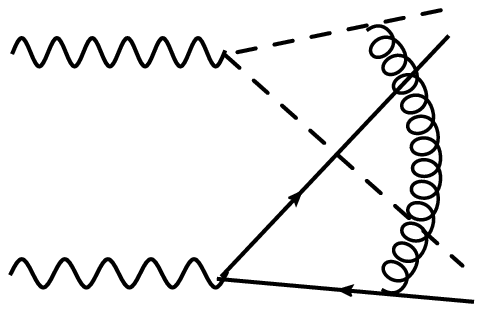}

   \caption{\label{fig:diagram2}
   \small Heavy meson pair production in photon-photon collision.
The wiggly (curly) line is for a photon (gluon), 
while the solid (dotted) one is for the heavier (lighter) (anti)quark.
}
\end{figure}

In this section we discuss how to calculate the elementary cross section for the $\gamma \gamma \to D \bar D$ reaction.
For the rather heavy mesons a pQCD approach seems the best method to calculate the cross section.
In the leading order of $\alpha_s$, 20 Feynman diagrams are involved
as shown in Fig.\ref{fig:diagram2}.
The diagrams can be classified into three groups.
Six diagrams of first part in Fig.\ref{fig:diagram2} represent the heavier- quark pair
production by two- photon collisions followed by one virtual
gluon emission to allow the produced heavier quarks to
hadronize into heavy mesons.
The second part of the figures shows diagrams obtained by exchanging heavier quark lines 
with lighter antiquark lines.
The last part consists of eight diagrams where
one photon produces a pair of heavier quarks and the other photon produces a pair of lighter quarks.

In Ref.\cite{Choi1994} the distribution amplitudes of the form:
\begin{equation}
\Phi_M(x,Q^2) = f(x,Q^2) \frac{1 + V}{2} \gamma_5, 
\Phi_{\bar M}(x,Q^2) = f^*(x,Q^2) \frac{1 - V}{2} \gamma_5 \;
\end{equation}
have been assumed with the heavy-quark approximation
\begin{equation}
f(x,Q^2) = f^*(x,Q^2) = \delta(x - \Lambda/M) \; ,
\end{equation}
where $\Lambda = M - m_Q$.

The resulting pseudoscalar-pseudoscalar (PP) production amplitude $M_{PP}^{\gamma\gamma}(\lambda,\lambda')$ can be written as \cite{Choi1994}

\begin{eqnarray}
M_{PP}^{\gamma\gamma}(\lambda, \lambda')=
2\frac{F^{\gamma \gamma}}{(1-z^2)^2}
(1+z^2)\Big [e_Q^2 F_{PP} (\lambda,\lambda')
+e_q^2 F'_{PP} (\lambda,\lambda')\Big ]
-2 e_Q e_q G_{PP} (\lambda, \lambda')
\nonumber
\end{eqnarray}

\begin{equation}
-(1-z^2)\Big [\frac{1-x}{x} e_Q^2 H_{PP} (\lambda,\lambda')
+ \frac {x}{1-x} e_q^2 H'_{PP} (\lambda,\lambda')\Big ],
\end{equation}

where
\begin{eqnarray}
F^{\gamma \gamma}= 
\frac{16\pi^2\alpha_e\alpha_sC_F}{{\hat s}x^2(1-x)^2}
\left[ \frac{f_M}{M} \right],
\end{eqnarray}

\begin{eqnarray}
F_{PP} (\lambda,\lambda')=
\Big [(1-x)[2-x(\hat s +2)]+ \frac{\hat s}{2}\sigma_{\lambda,-\lambda'}\Big ](\beta^2-z^2),
\end{eqnarray}

\begin{eqnarray}
F'_{PP} (\lambda,\lambda')=
\Big [x-[2-(1-x)(\hat s +2)]\frac{\hat s}{2}\sigma_{\lambda,-\lambda'}\Big ](\beta^2-z^2),
\end{eqnarray}

\begin{eqnarray}
G_{PP} (\lambda,\lambda')=
\Big [\frac{\hat s}{2}[1+z^2+(1-z^2)\sigma_{\lambda,-\lambda'}]-2x(1-x(2+\hat s))\Big ]\\
\nonumber
(\beta^2-z^2)-[\hat s(1-z^2)-2(3-z^2)\sigma_{\lambda,-\lambda'},
\end{eqnarray}

\begin{eqnarray}
H_{PP} (\lambda,\lambda')=
[2-x(\hat s +2)]\Big [(1-x)\beta^2-z^2)+z^2\sigma_{\lambda,-\lambda'}\Big ]+x(\hat s+2)
\sigma_{\lambda,-\lambda'},
\end{eqnarray}

\begin{eqnarray}
H'_{PP} (\lambda,\lambda')=
\Big [2-(1-x)(\hat s +2)\Big ]\Big [x(\beta^2-z^2)+z^2\sigma_{\lambda,-\lambda'}\Big]\\
\nonumber
+(1-x)+x(\hat s+2)
\sigma_{\lambda,-\lambda'}.
\end{eqnarray}
In the formulas above:
$\hat s= \frac{s}{M^2}$, $z=\beta cos\theta$, and $\beta= \sqrt{ 1-\frac{4}{\hat s}}$.\\
The $\theta$ is the scattering angle between the photon and a heavy meson, 
the color factor $C_F=\frac{4}{3}$.

In the $e^+e^- \to e^+e^-\pi^+\pi^-$ reaction usually cuts on $z= cos\theta$, where $\theta$ is the angle in the photon-photon center of mass frame, are imposed.
This type of cuts is rather difficult to realize in the $AA \to AAD\bar D$ reactions where kinematics is usually incomplete. However, very often transverse momenta of charged pions can be easily measured with modern detectors. Therefore in the following we impose cuts on transverse momenta of both charmed mesons. In order not to loose too much of the signal we require only $p_t >$ 1 GeV.

\begin{figure}[!thb]
\begin{center}
\includegraphics[width=6.3cm]{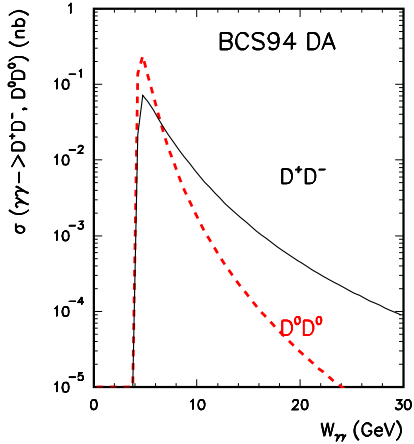}
\caption{
\label{fig:sigma}
The energy dependence of the $\gamma \gamma \to D^+D^-$ (black line)
and $\gamma \gamma \to D^0 \bar{D^0}$ (dashed, red online) in the heavy-quark approximation
\cite{Choi1994}. In this calculations we have imposed an extra cut on the D-meson
transverse momenta $p_t >$ 1 GeV.
}
\end{center}
\end{figure}

In Fig.\ref{fig:sigma} we show the energy dependence of the elementary 
$\gamma \gamma \to D^+D^-(D^0 \bar{D^0})$ cross sections calculated according to
the formalism used in Ref. \cite{Choi1994} with the cut on meson transverse momenta.
The cross section for $D^0 \bar {D^0}$ is larger than that for $D^+D^-$ close to the  threshold but falls faster with increasing $W_{\gamma\gamma}$.


Above we have used the heavy quark approximation, i.e. we have assumed that light
quark/antiquark carries a fixed fraction of the final meson momentum,
$x = \frac{\Lambda}{M} = \frac{M-m_a}{M}$
and used hard process matrix elements with finite heavy quark masses.
At present, it is known that the distribution amplitude is not of the 
Dirac delta type (see e.g. \cite{Huang} and references therein).\\
Therefore in the following we use also the classical Brodsky-Lepage formalism
\cite{BL81} with distribution amplitudes taken from the recent studies
 \cite{Huang}. 
In this approach the normalized to unity distribution amplitude is given as
\begin{equation}
\phi_D(x)=\frac{\sqrt{6}A_Dy}{8\pi^{3/2}f_D}\sqrt{x(1-x)}\Big [1-Erf(\frac{b_Dy}{\sqrt{x(1-x)}})\Big ]\exp{\left[-b_D^2\frac{(xm_2^{*2}+(1-x)m_1^{*2}-y^2)}{x(1-x)}\right]},
\label{WH_amplitude}
\end{equation}
where $y=x m_2^*+(1-x)m_1^*$, $Erf(x)=\frac{2}{\pi}\int^x_0{\exp({-t^2})dt}$ and
$m_1^*$ and $m_2^*$ are quark (constituent) masses.
We take the parameters in Eq.(\ref{WH_amplitude}) from \cite{Huang}:
$P_D\simeq0.8$, $A_D=116~\mbox{GeV},~~~b_D=0.592~\mbox{GeV}^{-1}$,  $f_D=0.223~\mbox{GeV}$.

In the Brodsky-Lepage formalism the amplitude for the $\gamma \gamma \to D \bar D$
process can be written schematically as:
\begin{equation}
{\cal M}_{\gamma \gamma \to D \bar D} = \int \Phi^*(x,\mu^2)  H_{\lambda_{1} \lambda_{2}}(x,y,z) \Phi^*(y,\mu^2) \; dx dy \; .
\label{BL_amplitude}
\end{equation}
Above $H_{\lambda_{1} \lambda_{2}}(x,y,z)$ is the photon helicity dependent hard matrix element for $\gamma \gamma \to q \bar Q \bar q Q$
and $\Phi$'s are Brodsky-Lepage distribution amplitudes.
The distribution amplitude $\Phi$ here is related to the distribution amplitude in 
Eq.(\ref{WH_amplitude}) as:
\begin{equation}
\Phi(x) = \frac{f_D}{2\sqrt{3}} \phi_D(x).
\end{equation}

\begin{figure}[!thb] 
\begin{center}
\includegraphics[width=6.3cm]{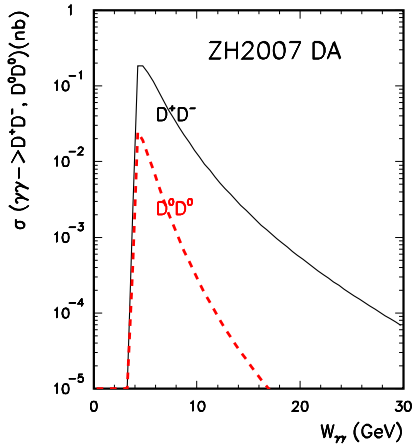}
\caption{
\label{fig:sigma_W_BL}
The energy dependence of the $\gamma \gamma \to D^+D^-$ (black line)
and $\gamma \gamma \to D^0 \bar {D^0}$ (dashed, red online) in the Brodsky-Lepage formalism
\cite{BL81} with distribution amplitude from \cite{Huang}.
In this calculations we have imposed an extra cut on the D-meson
transverse momenta ($p_t >$ 1 GeV).
}
\end{center}
\end{figure}

In Fig.\ref{fig:sigma_W_BL} we show elementary cross section as a function of the $\gamma \gamma$
subsystem energy for the Brodsky-Lepage formalism. As in the heavy- quark approximation an extra cut on the
$D$ meson transverse momentum, $p_t >$ 1 GeV, has been imposed.
While the cross section for $D^+D^-$ is similar in both approaches, the cross sections for $D^0 \bar {D^0}$ differ considerably.

\section{Nuclear collisions, Equivalent Photon Approximation}

\label{section:nuclear}

To calculate the cross section of the nuclear process it is convenient to introduce a new kinematic variable: 
$x = \frac{\omega}{E_A}$, where $\omega$ is the energy of the photon and the energy of the nucleus 
$E_A=\gamma A m_{proton} = \gamma M_A$, where $M_A$ is the mass of the nucleus 
and $\gamma$ is the Lorentz factor.

The total cross section can be calculated by the convolution \cite{KS2010}:
\begin{eqnarray}
 \sigma\left(AA \rightarrow AA D \bar D ;s_{AA}\right) = 
\int   {\hat \sigma} \left(\gamma\gamma\rightarrow D \bar D;
W_{\gamma \gamma} = \sqrt{x_1 x_2 s_{AA}} \right) \, 
{\rm d}n_{\gamma\gamma}\left(x_1,x_2,{\bf b}\right)  .
\end{eqnarray}
The effective photon fluxes can be expressed through the electric 
fields generated by the nuclei:
\begin{eqnarray}
 {\rm d}n_{\gamma\gamma}\left(x_1,x_2,{\bf b}\right) = &  & \frac{1}{\pi} {\rm d^2}{\bf b}_1 |{\bf E}\left(x_1,{\bf b}_1 \right)|^2 \frac{1}{\pi} {\rm d^2}{\bf b}_2  |{\bf E}\left(x_2,{\bf b}_2 \right)|^2   \nonumber \\
 & \times & S^2_{abs}\left({\bf b} \right) \delta^{\left(2\right)} \left({\bf b}-{\bf b}_1 + {\bf b}_2 \right)   \frac{{\rm d}x_1}{x_1} \frac{{\rm d}x_2}{x_2}. 
\end{eqnarray}
The presence of the absorption factor $S^2_{abs}\left({\bf b} \right)$ 
assures that we consider only peripheral collisions, when the nuclei 
do not undergo nuclear breakup. In the first approximation this can be
expressed as:  
\begin{eqnarray}
 S^2_{abs}\left({\bf b} \right)=\theta \left({\bf b}-2R_A \right) = \theta \left(|{\bf b}_1-{\bf b}_2|-2R_A \right) \; .
\end{eqnarray}
Thus in the present case, we concentrate on processes with final 
nuclei in the ground state. 
The electric field strength can be expressed through the charge 
form factor of the nucleus:
\begin{eqnarray}
 {\bf E}\left(x,{\bf b}\right) = Z \sqrt{4\pi \alpha_{em}} \int \frac{{\rm d^2}{\bf q}} {\left(2\pi^2\right)} e^{-i{\bf bq}}\frac{{\bf q}}{{\bf q}^2+x^2M_A^2}F_{em}\left({\bf q}^2+x^2M_A^2\right).
\end{eqnarray}
Next we can benefit from the following formal substitution:
\begin{eqnarray}
\frac{1}{\pi} \int {\rm d^2}{\bf b} |{\bf E}\left(x,{\bf b}\right)|^2 
= \int {\rm d^2}{\bf b} N \left( \omega, {\bf b} \right) 
\equiv n\left(\omega \right) \; 
\end{eqnarray}
by introducting effective photon fluxes which depend on energy of 
the quasireal photon $\omega$ and the distance from the nucleus 
in the plane perpendicular to the nucleus motion $\overrightarrow{b}$. 
Then, the luminosity function can be expressed in term of the photon flux 
factors attributed to each of the nuclei
\begin{eqnarray}
 {\rm d}n_{\gamma\gamma}\left(\omega_1,\omega_2,{\bf b}\right) & = & \int  \theta \left(|{\bf b}_1-{\bf b}_2|-2R_A \right) N \left(\omega_1,{\bf b}_1 \right)    N\left(\omega_2,{\bf b}_2 \right) {\rm d^2}{\bf b}_1 {\rm d^2}{\bf b}_2 \frac{{\rm d}\omega_1}{\omega_1} \frac{{\rm d}\omega_2}{\omega_2}. 
\end{eqnarray}
The total cross section for the $AA \rightarrow AA D \bar D$ 
process can be factorized into the equivalent photons spectra
( $n \left( \omega \right)$ ) and the 
$\gamma \gamma \to D \bar D$ subprocess cross section as 
(see e.g.\cite{BF91}):
\begin{eqnarray}
 \sigma\left(AA \rightarrow AA D \bar D ; s_{AA}\right)  & = &  
\int   
{\hat \sigma}\left(\gamma\gamma\rightarrow D \bar D; 
W_{\gamma \gamma}  \right) \, 
\theta \left(|{\bf b}_1-{\bf b}_2|-2R_A \right)  \nonumber \\
 &\times & N\left(\omega_1,{\bf b}_1 \right)
N\left(\omega_2,{\bf b}_2 \right)
  {\rm d^2}{\bf b}_1 
 {\rm d^2}{\bf b}_2   \frac{{\rm d}\omega_1}{\omega_1} \frac{{\rm d}\omega_2}{\omega_2} \; ,
\label{eq.tot_cross_section}
\end{eqnarray}
where $W_{\gamma \gamma}=\sqrt{4 \omega_1 \omega_2}$ is energy in the $\gamma \gamma$ subsystem.

\begin{equation}
\sigma \left( AA \rightarrow AA D \bar D  \right) 
= \int {\hat \sigma} \left(\gamma\gamma\rightarrow D \bar D; 
\sqrt{4 \omega_1 \omega_2} \right) n\left( \omega_1 \right) 
 n\left( \omega_2 \right) \frac{d \omega_1}{\omega_1}
\frac{d \omega_2}{\omega_2}  \; .
\end{equation}
Additionally, we define 
$ Y=\frac{1}{2} \left( y_{D}+y_{\bar D}\right)$, rapidity of 
the outgoing meson system which is produced in the photon--photon 
collision. 
Performing the following transformations:
\begin{equation}
\omega_1 = \frac{W_{\gamma \gamma}}{2}e^Y, \qquad \omega_2 = \frac{W_{\gamma \gamma}}{2}e^{-Y} \; ,
\label{eq:omega}
\end{equation}
\begin{equation}
\frac {{\rm d}\omega_1} {\omega_1} \frac{{\rm d}\omega_2} {\omega_2} = \frac{2}{W_{\gamma \gamma}} {\rm d}W_{\gamma \gamma} {\rm d} Y \; ,
\label{eq:transf}
\end{equation}
\begin{equation}
{\rm d} \omega_1 {\rm d} \omega_2 \to {\rm d} W_{\gamma \gamma} {\rm d} Y \mbox{ where } \left|\frac{\partial \left( \omega_1, \omega_2 \right) }{\partial \left( W_{\gamma \gamma}, Y \right)}  \right| = \frac{W_{\gamma \gamma }}{2}
\; ,
\label{eq:transf_jac}
\end{equation}
formula (\ref{eq.tot_cross_section}) can be rewritten as:
\begin{eqnarray}
 \sigma\left(AA \rightarrow  AA D \bar D ; s_{AA}\right)  = 
 \int   {\hat \sigma}\left(\gamma\gamma\rightarrow D \bar D; W_{\gamma \gamma}  \right) \theta \left(|{\bf b}_1-{\bf b}_2|-2R_A \right) & & \nonumber \\
 \times N\left(\omega_1,{\bf b}_1 \right) 
 N\left(\omega_2,{\bf b}_2 \right)
   \frac{2}{W_{\gamma \gamma}} {\rm d^2}{\bf b}_1  
 {\rm d^2}{\bf b}_2   
 {\rm d}W_{\gamma \gamma} {\rm d} Y
& \; & .
 \label{eq.tot_cross_section_WY}
\end{eqnarray}
Finally, the nuclear cross section can be expressed as the five-fold 
integral:
\begin{eqnarray}
 \sigma \left(AA  \rightarrow   AA D \bar D ; s_{AA}\right)  = 
 \int  {\hat \sigma}\left(\gamma\gamma\rightarrow D \bar D; W_{\gamma \gamma}  \right) \theta \left(|{\bf b}_1-{\bf b}_2|-2R_A \right)& & \nonumber \\ 
   \times   N \left(\omega_1,{\bf b}_1 \right) N\left(\omega_2,{\bf b}_2 \right)2 \pi b_m \, {\rm d} b_m \, {\rm d} \overline{b}_x \, {\rm d} \overline{b}_y \frac{W_{\gamma \gamma}}{2} {\rm d}W_{\gamma \gamma} {\rm d} Y & \, & , 
\label{eq.tot_cross_section_our}
\end{eqnarray}
where $\overline{b}_x \equiv (b_{1x}+b_{2x})/2$,
      $\overline{b}_y \equiv (b_{1y}+b_{2y})/2$ and
$\vec{b}_m = \vec{b}_1 - \vec{b}_2$ have been introduced.
This formula is used to calculate the total cross section
for the $A A \to A A D \bar D$ reaction as well as the distributions
in the impact parameter $b = b_m$, the meson invariant mass $W_{\gamma \gamma} = M_{D \bar D}$ and the meson pair rapidity $Y(D \bar D)$.

\section{Results for the nuclear collisions}

In this section we present results
for gold-gold collisions at RHIC ($W_{NN}$ = 200 GeV) and lead-lead collisions at LHC 
($W_{NN}$ = 5.5 TeV). In all the calculations presented here
we use realistic charge form factor being a Fourier transform of the realistic charge densities.
Both $D^+ D^-$ and $D^0 \bar {D^0}$ channels are discussed.
Below we present quantities which can be easily calculated in the b-space EPA discussed in the previous
section.
In Table I we have collected numerical values of the cross sections for considered exclusive 
processes. The cross sections are small but the reactions seem measurable especially at LHC.

\begin{figure}[!thb] 
\begin{center}
\includegraphics[width=6.3cm]{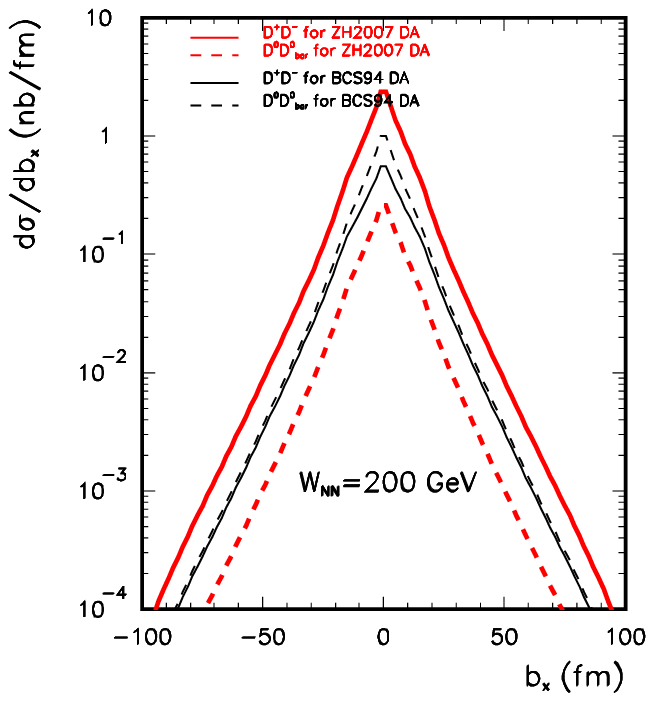}
\includegraphics[width=6.3cm]{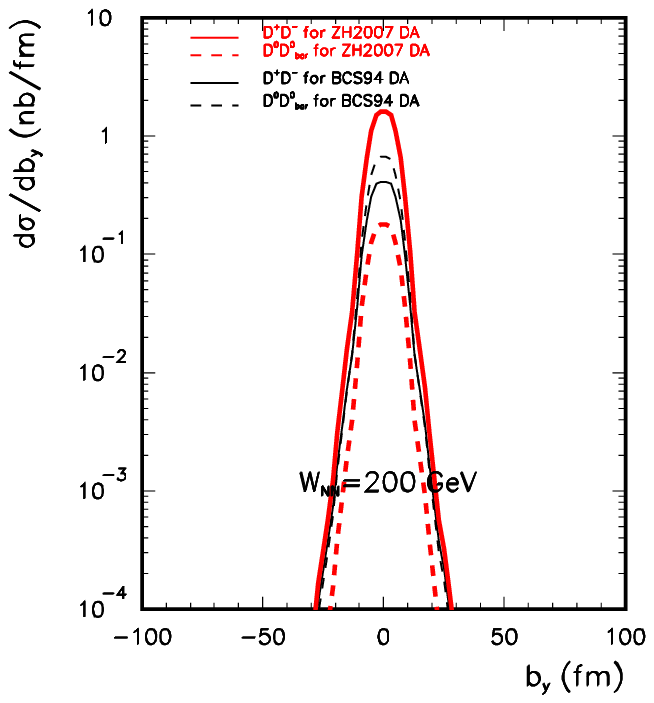}
\includegraphics[width=6.3cm]{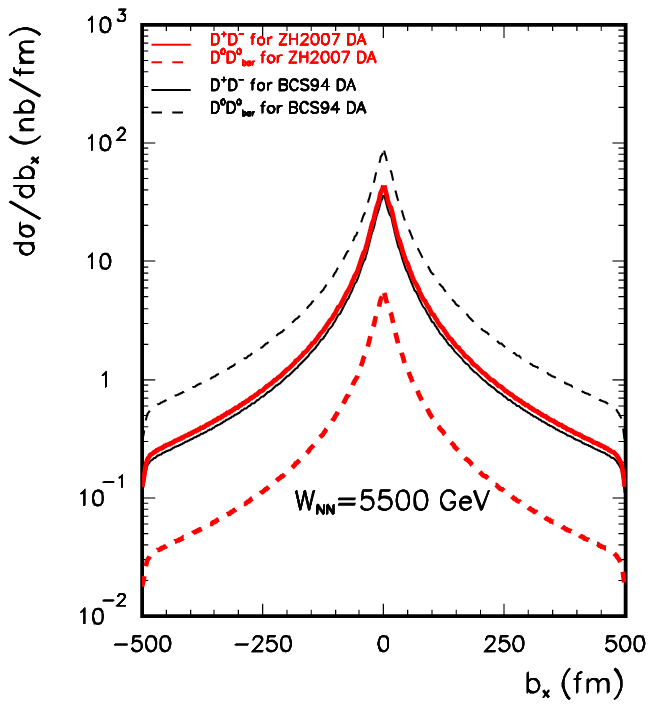}
\includegraphics[width=6.3cm]{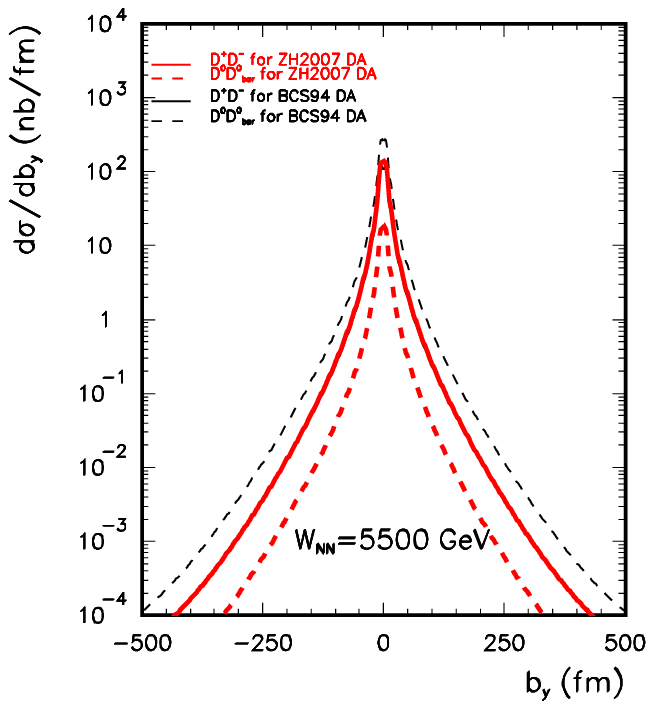}
\caption[*]{
\label{fig:dsig_dbxy}
The distributions in the auxiliary quantities $b_x$ and $b_y$ defined in section \ref{section:nuclear}
for $D^+ D^-$ and $D^0 {\bar D}^0$ for the BCS94 DA \cite{Choi1994} (black) and for the ZH2007 DA
\cite{Huang} (red) for RHIC and LHC
energies.
}
\end{center}
\end{figure}

Before we go to the presentation of more interesting quantities we wish to show a control plot.
In Fig.\ref{fig:dsig_dbxy} we show distributions in the auxiliary variables $b_x$ and $b_y$
defined in Section \ref{section:nuclear}. These distributions are peaked at
$b_x, b_y$ = 0 and strongly decrease with increasing $|b_x|$ or $|b_y|$.
This figure demonstrates that the chosen range of integration over the auxiliary
variables $b_x$ and $b_y$ is sufficient. Please note that the range of integration for 
the LHC energy is 5 times larger than that for the RHIC energy.
To speed up calculation we have performed integration
in only two quadrants in the $(b_x, b_y)$ space and used relevant symmetry relations.

\begin{table}

\caption{Total cross section for the exclusive $D \bar D$ production for RHIC ($s_{NN}^{1/2}=200$ GeV) and LHC
($s_{NN}^{1/2}=5500$ GeV)
 calculated with distribution amplitudes from \cite{Choi1994} and \cite{Huang}.}

\begin{tabular}{|c||c|c|c|} \hline

  Process                               & \multicolumn{2}{|c|}{$\sigma_{tot}$} \\
                                        & BCS94 DA             & ZH2007 DA      \\ \hline \hline

 $AuAu\to AuAu \, D^+ D^-$ & 12.44 $nb$      & 0.08 $\mu b$   \\ \hline
 $AuAu\to AuAu \, D^0 \bar D^0$ & 21.2 $nb$    & 5.06 $nb$      \\ \hline
\hline
 $PbPb\to PbPb \, D^+ D^-$ & 1.68 $\mu b$      & 1.92 $\mu b$   \\ \hline
 $PbPb\to PbPb \, D^0 \bar D^0$ & 4.09 $\mu b$    & 0.28 $\mu b$      \\ \hline

\end{tabular}

\end{table}


\begin{figure}[!thb] 
\begin{center}
\includegraphics[width=6.3cm]{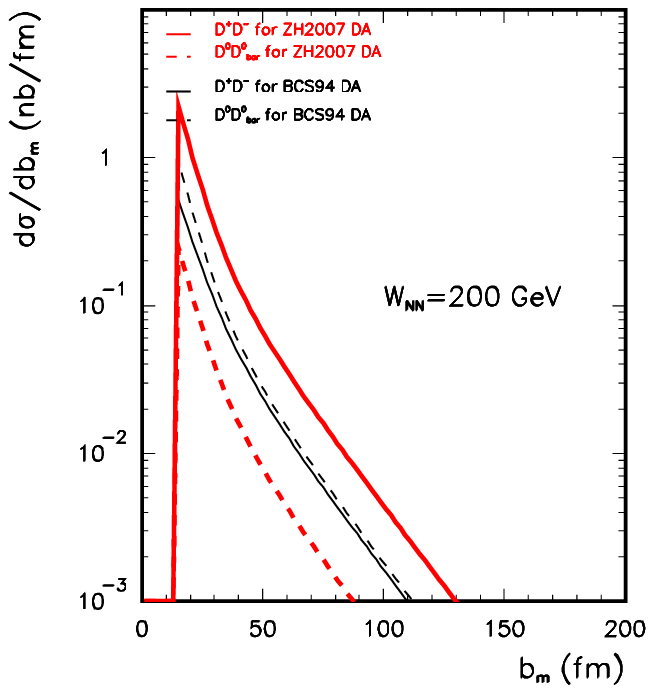}
\includegraphics[width=6.3cm]{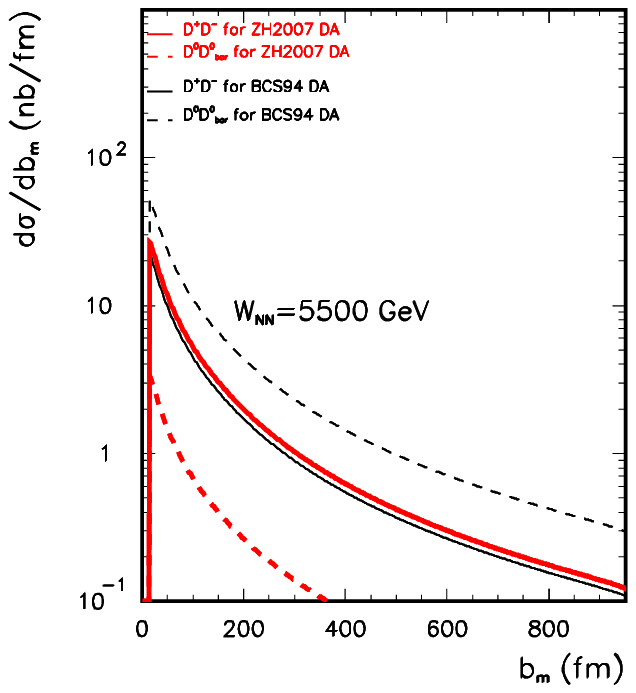}
\caption[*]{
Distribution in impact parameter for $D^+ D^-$
and $D^0 {\bar D}^0$  for the BCS94 DA \cite{Choi1994} (black) and for the ZH2007 DA
\cite{Huang} (red) for RHIC and LHC energies.
\label{fig:dsig_dbm}
}
\end{center}
\end{figure}

In the heavy-ion photon-photon processes particles can be produced
when nuclei fly far away apart. This is particularly spectacular for
light leptons (see e.g.\cite{KS2010}). In Fig.\ref{fig:dsig_dbm}
we show such a distribution for the pseudoscalar $D \bar D$ production.
The maximum of the cross section is for the impact parameter $b$ when colliding nuclei almost touch each other and the cross section decrease for larger $b$.
The decrease is much sharper for RHIC than for LHC. Compared to light leptons \cite{KS2010}
here it is easier to achieve a convergence in $b$.


\begin{figure}[!thb] 
\begin{center}
\includegraphics[width=6.3cm]{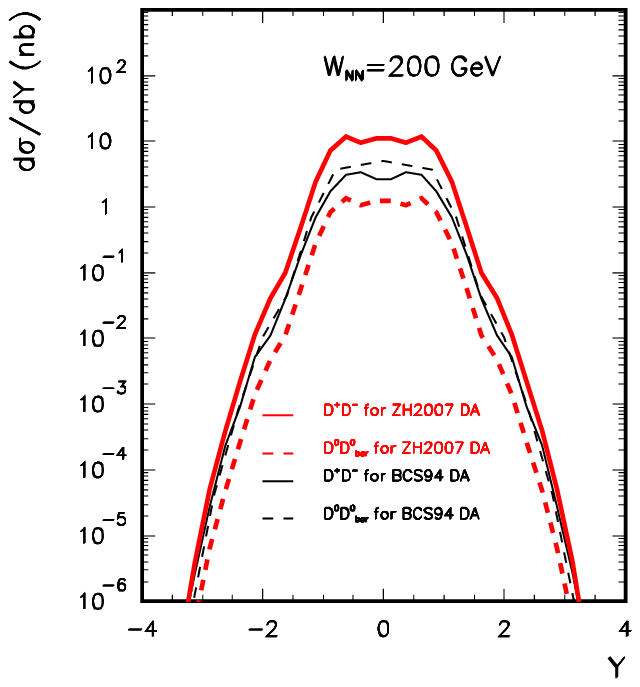}
\includegraphics[width=6.3cm]{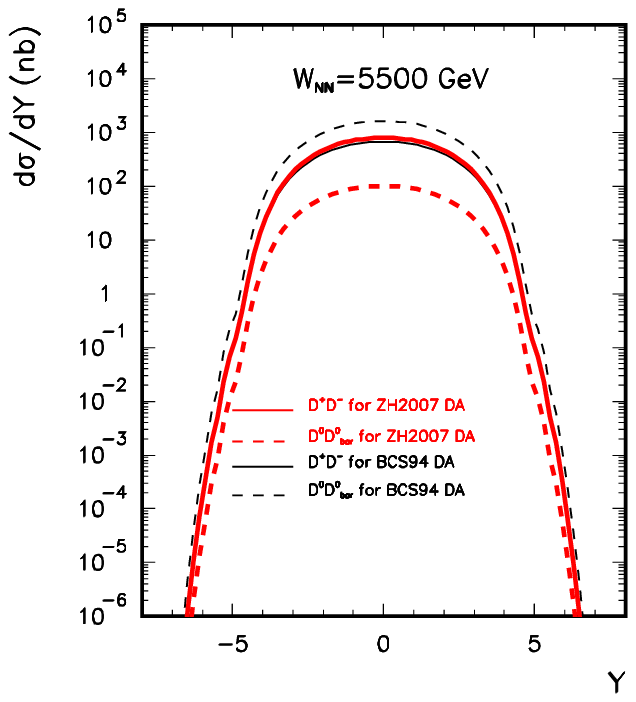}
\caption{
\label{fig:dsig_dy}
The cross section as a function of 
the meson pair rapidity for $D^+ D^-$ and $D^0  {\bar D}^0$ for the BCS94 DA \cite{Choi1994} (black) and for the ZH2007 DA \cite{Huang} (red) for RHIC and LHC energies.
}
\end{center}
\end{figure}

Now we come to the distributions which can be directly measured in experiments.
In Fig.\ref{fig:dsig_dy} we show the distribution in the $D^+D^-$ and $D^0 {\bar D}^0$ pair 
rapidity being $Y \approx \frac{1}{2} (y_{D} + y_{\bar D})$. The visible irregularities at larger $|Y|$
are caused by the oscillating nuclear form factor. The larger meson pair rapidity the larger four-momentum squared
of the exchanged photon (which is the argument of the charge form factor).
The irregularities correspond to the four-momentum squared when the charge form factor changes its sign, i.e. when $|F|^2$ is close to zero.
A more elaborate discussion on the role of the nuclear form factor can be found in Ref.\cite{KS2010}.


\begin{figure}[!thb]
\begin{center}
\includegraphics[width=6.3cm]{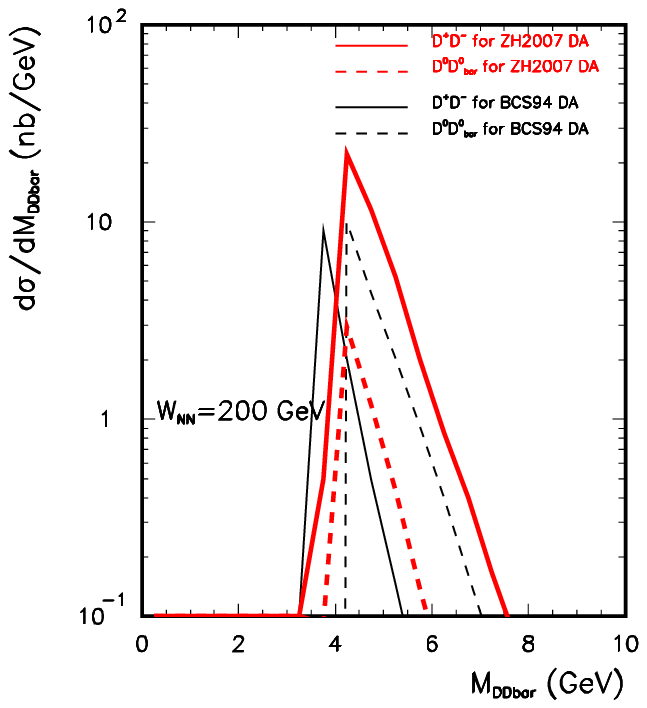}
\includegraphics[width=6.3cm]{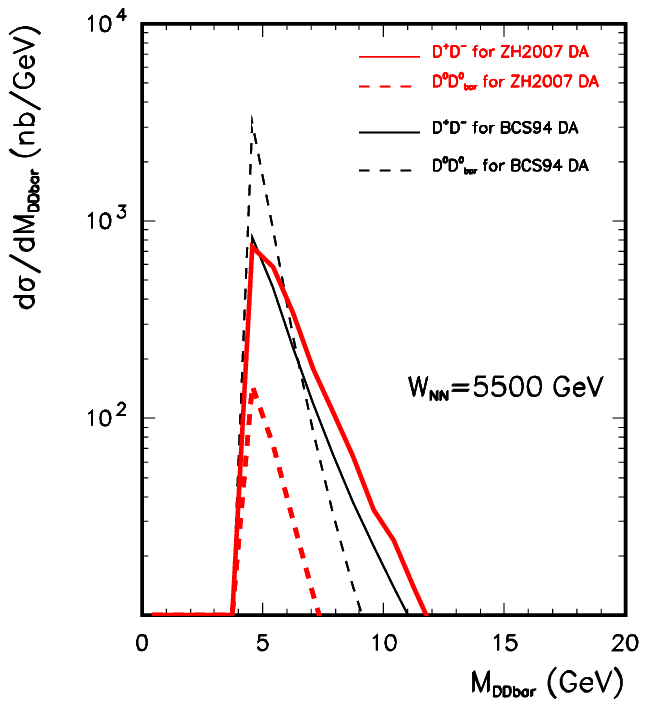}
\caption[*]{$D \bar D$ invariant mass distribution
for RHIC and LHC.
\label{fig:dsig_dMDD}
}
\end{center}
\end{figure}

In Fig.\ref{fig:dsig_dMDD} we show distribution in invariant mass
of $D$ and $\bar D$. We predict steep fall-off of the disribution as a function
of the invariant mass, even steeper than for the elementary cross section shown in Fig.\ref{fig:sigma}. This figure shows that in experiments only a region of
small invariant masses close to the threshold could be investigated.


\section{Conclusions}
We have calculated for the first time in the literature total and differential cross sections for exclusive production of pseudoscalar $D \bar D$ meson pairs in the $AA \to AAD\bar D$ reaction assuming
that the reaction is driven by the 
$\gamma \gamma \to D \bar D$ subproceess.
The elementary cross sections were calculated in the heavy - quark approach as well as in the Brodsky- Lapage formalism with distribution amplitude describing recent CLEO data on leptonic $D^+$ decay. Rather small cross sections have been found. The cross section for exclusive
 $D \bar D$ production is much smaller than the cross section for the exclusive or semi-exclusive production
of $c \bar c$ calculated recently. In our calculations absorption effects were included in the impact parameter Equivalent Photon Approximation.
The meson pairs are produce preferentially when the nuclei almost touch each other.
The cross section strongly depends on the approximation made in the calculation.
The dominant contribution to the cross section comes from the region of very small $D \bar D$
invariant masses.

Whether the process can be measured requires further dedicated studies including detector simulations. We belive that our evaluation will be a useful starting point for such future studies.
Since the cross section strongly depends on theoretical details experimental verification would be very helpful.

\vspace{1cm}

{\bf Acknowledgments}

We are indebted to Mariola K{\l}usek-Gawenda and Wolfgang Sch\"afer for a discussion
of some details of the present analysis.
A nice conversation with S.Y. Choi is acknowledged.
This work was partially supported by the Polish grant MNiSW N N202 249235.


%
%

%
\end{document}